\documentclass{icrc}

\usepackage{times}
\ifx\pdfoutput\undefined 
\usepackage{graphicx}
\else
\usepackage[pdftex]{graphicx}
\fi

\def\la{\mathrel{\mathchoice {\vcenter{\offinterlineskip\halign{\hfil
$\displaystyle##$\hfil\cr<\cr\sim\cr}}}
{\vcenter{\offinterlineskip\halign{\hfil$\textstyle##$\hfil\cr<\cr\sim\cr}}}
{\vcenter{\offinterlineskip\halign{\hfil$\scriptstyle##$\hfil\cr<\cr\sim\cr}}}
{\vcenter{\offinterlineskip\halign{\hfil$\scriptscriptstyle##$\hfil\cr<\cr
\sim\cr}}}}}

\def\ga{\mathrel{\mathchoice {\vcenter{\offinterlineskip\halign{\hfil
$\displaystyle##$\hfil\cr>\cr\sim\cr}}}
{\vcenter{\offinterlineskip\halign{\hfil$\textstyle##$\hfil\cr>\cr\sim\cr}}}
{\vcenter{\offinterlineskip\halign{\hfil$\scriptstyle##$\hfil\cr>\cr\sim\cr}}}
{\vcenter{\offinterlineskip\halign{\hfil$\scriptscriptstyle##$\hfil\cr>\cr
\sim\cr}}}}}

\begin{document}

\title{The Physics of the Knee in the Cosmic Ray Spectrum}

\author[1,2]{K-H.~Kampert}
\affil[1]{Institut f\"ur Experimentelle Kernphysik, University
              of Karlsruhe, 76021~Karlsruhe, Germany}
\affil[2]{Institut f\"ur Kernphysik, Forschungszentrum Karlsruhe, 
          76021~Karlsruhe, Germany}
\author[1]{T.~Antoni}
\author[2]{W.D.~Apel}
\author[1,*]{F.~Badea}
\author[2]{K.~Bekk}
\author[2,*]{A.~Bercuci}
\author[2,1]{H.~Bl\"umer}
\author[2]{E.~Bollmann}
\author[3]{H.~Bozdog}
\author[3]{I.M.~Brancus}
\author[2]{C.~B\"uttner}
\author[4]{A.~Chilingarian}
\author[1]{K.~Daumiller}
\author[2]{P.~Doll}
\author[2]{J.~Engler}
\author[1]{F.~Fe{\ss}ler}
\author[2]{H.J.~Gils}
\author[1]{R.~Glasstetter}
\author[1]{R.~Haeusler}
\author[2]{A.~Haungs}
\author[2]{D.~Heck}
\author[1]{J.R.~H\"orandel}
\author[2]{T.~Holst}
\author[1,5]{A.~Iwan}
\author[5,+]{J.~Kempa}
\author[2]{H.O.~Klages}
\author[1,$\P$]{J.~Knapp}
\author[2]{G.~Maier}
\author[2]{H.J.~Mathes}
\author[2]{H.J.~Mayer}
\author[1]{J.~Milke}
\author[2]{M.~M\"uller}
\author[2]{R.~Obenland}
\author[2]{J.~Oehlschl\"ager}
\author[1]{S.~Ostapchenko}
\author[3]{M.~Petcu}
\author[2]{H.~Rebel}
\author[2]{M.~Risse}
\author[2]{M.~Roth}
\author[2]{H.~Schieler}
\author[2]{J.~Scholz}
\author[2]{T.~Thouw}
\author[1]{H.~Ulrich}
\author[3]{B.~Vulpescu}
\author[1]{J.H.~Weber}
\author[2]{J.~Wentz}
\author[2]{J.~Wochele}
\author[6]{J.~Zabierowski}
\author[2]{S.~Zagromski}
\affil[3]{National Institute of Physics and Nuclear Engineering,
              7690~Bucharest, Romania}
\affil[4]{Cosmic Ray Division, Yerevan Physics Institute,
              Yerevan~36, Armenia}
\affil[5]{Department of Experimental Physics,
              University of Lodz, 90236~Lodz, Poland}
\affil[6]{Soltan Institute for Nuclear Studies,
              90950~Lodz, Poland}
\affil[$\P$]{\small now at: University of Leeds, Leeds LS2 9JT, U.K.}
\affil[$*$]{\small on leave of absence from IFIN-HH, Bucharest}
\affil[+]{\small now at: Warsaw University of Technology, 
                 09-400~Plock, Poland}

\correspondence{K.-H. Kampert (kampert@ik.fzk.de)}

\firstpage{1}
\pubyear{2001}

\titleheight{110mm}

\maketitle

\begin{abstract}
~~Recent results from the KASCADE extensive air shower experiment
are presented.  After briefly reviewing the status of the
experiment we report on tests of hadronic interaction models 
and emphasize the progress being made in understanding
the properties and origin of the knee at $E_{knee}\cong
4\cdot10^{15}$ eV.\\
Analysing the muon- and hadron trigger rates in the KASCADE 
calorimeter as well as the global properties of high energy hadrons
in the shower core leads us to conclude that
QGSJET still provides the best overall description of EAS data, 
being superior to DPMJET II-5 and {\sc neXus 2}, for example.\\
Performing high statistics CORSIKA simulations and applying
sophisticated unfolding techniques to the electron and muon
shower size distributions, we are able to successfully
deconvolute the all-particle energy spectrum into energy spectra
of 4 individual primary mass groups (p, He, C, Fe).  Each of
these preliminary energy distributions exhibits a knee like
structure with a change of their knee positions suggesting a
constant rigidity of $R \cong 2$-3 PV.
\end{abstract}

\section{Introduction}

The origin and acceleration mechanism of ultra-high energy cosmic
rays ($E \ga 10^{14}$ eV) have been subject to debate for several
decades.  Mainly for reasons of the power required to maintain
the observed cosmic ray energy density of $\varepsilon_{\rm cr}
\approx 1$ eV/cm$^{3}$, the dominant acceleration sites are
generally believed to be supernova remnants (SNR).  Charged
particles mainly originating from the surrounding interstellar
medium of the pre-supernova star may get trapped at the highly
supersonic shock wave generated by the SN explosion.  Repeatedly
reflections on both sides of the shock front lead to an
acceleration by the so-called `first order Fermi mechanism'. 
Naturally, this leads to a power law spectrum $dJ/dE \propto
E^{-\gamma}$ as is observed experimentally.  Simple dimensional
estimates show that this process is limited to $E_{\rm max} \la Z
\times (\rho \times B)$, with $Z$ being the atomic number of the
cosmic ray (CR) isotope and $\rho$, $B$ the size and magnetic
field strength of the acceleration region.  A more detailed
examination of the astrophysical parameters suggests an upper
limit of acceleration of $E_{\rm max} \approx Z\times
10^{15}$\,eV \citep{drury94b,berezhko99}.  Curiously, the CR
spectrum steepens from $\gamma \simeq 2.75$ to $\simeq 3.1$ at $E
\simeq 4 \times 10^{15}$\,eV which is called the `knee'.  The
coincidence thus may indicate that the `knee' is related to the
upper limit of acceleration.

Alternative interpretations of the knee discuss a change in the
propagation of CRs from their sources to the solar system.  Such
kind of propagation effects are conveniently described by an
`escape time' from our galaxy, $\tau_{\rm esc}$.  Extrapolation
of $\tau_{\rm esc}$ from direct measurements in the GeV range to
higher energies breaks down at $E \approx 3\cdot10^{15}$ eV,
because $c\tau_{\rm esc} \sim 300$ pc which is the thickness of
the galactic disk \citep{gaisser00a}.  The value corresponds to
just one crossing of the disk and would give rise to significant
anisotropies with respect to the galactic plane when approaching
this value.  Similarly as to the process of acceleration at SNR
shocks, the process of galactic containment is closely related to
magnetic field confinement, i.e.\ in addition to anisotropies one
again expects $E_{\rm max}^{\rm gal} \propto Z$.

A picture related to both of these interpretations has been
proposed by \citet{erlykin97a}.  They consider the knee as a
superposition of a weakly energy dependent galactic modulation
with additional prominent structures in the flux spectrum caused
by a single near-by object.  This so-called 'single source model'
assumes that a shock wave of a recent nearby supernova which
exploded some 10,000 years ago at a distance of a few hundred
parsecs, currently propagates (or has recently propagated)
through the solar system causing distinct peaks of elemental
groups in the energy spectrum.  The most recent update of this
model can be found in the proceedings to this conference (p1804). 
However, there is some controversy whether or not the statistics
of presently existing data gives sufficient support to the model.

Besides those kind of astrophysical interpretations, also
particle physics motivated pictures of explaining the knee were
put forward.  For example, \citet{wigmans00} suggested the
inverse $\beta$-decay reaction $p + \bar{\nu_{e}} \to n + e^{+}$
with massive relic neutrinos could destroy protons.  Simple
kinematics shows that this channel is open for $E_{p} >
1.7\cdot10^{15} {\rm eV}/m_{\nu} {\rm (eV)}$.  Thus, a knee
energy $E_{\rm knee} \simeq 4$ PeV would correspond to an
electron neutrino mass of $m_{\nu} \sim$ 0.4 eV, a value
presently not excluded by any other observation or experiment. 
However, `eating' sufficiently large amounts of protons by such a
process requires extraordinary high local densities of relic
neutrinos, which appears doubtful even if possible gravitational
trapping is considered.  In case of a non-zero magnetic dipole
moment of massive neutrinos, also electromagnetic interactions
with much larger cross section would be possible, thus giving
rise to inelastic GZK-like $p\nu$-interactions.  Such a picture
requiring neutrinos masses of about 100 eV has recently been
discussed by \citet{dova-01}.

\citet{nikolsky95} suggested that the knee is not a property of
the primary energy spectrum itself, but may be caused by
changing high-energy interactions in the Earth's atmosphere. 
Producing a new type of a heavy particle in the first
interactions escaping unseen by air shower experiments, or an
abrupt increase in the multiplicity of produced particles (see
proceedings to this conference, p1389) could, in principle,
mimic a break in the spectrum.  From the particle physics point
of view this is not completely ruled out as the centre-of-mass
(cms) energy available at the knee is above Tevatron energies. 
However, a well worked out model simultaneously describing
accelerator and CR data without violating general physics, like 
the unitarity principle, is still to be presented.

Very recently, exotic scenarios with extra dimensions and
TeV-scale quantum gravity have been discussed by
\citet{kazanas01} and others.  Again, the argument is that the
cms-energy of the knee ($\sqrt{s}\cong 2$ TeV) is just above the
highest energies reached by present accelerators and that some
part of the CR energies is transferred into invisible channels, in
this case gravitational energy.  Some advantage of this picture is
that the required fast growth of cross section for graviton
radiation does not necessarily violate S-wave unitarity.

A common feature of the particle physics interpretations of the
knee is the expectation of seeing the break of different elements
to be displaced by their mass number $A$ rather than by their
nuclear charge $Z$.  This is understood from the reaction
mechanism being governed, at sufficiently high energy, by the
energy per nucleon $E/A$ of the incident particle.

To distinguish between these various ideas, better measurements
of the energy spectrum and elemental composition of cosmic rays
in the knee region are essential.  Most commonly, the composition
is expressed in terms of the mean logarithmic mass, $\langle \ln
A \rangle$.  Obviously, such an average quantity appears to be
rather insensitive to details of the individual energy spectra. 
Clearly, the most sensitive testbed for comparing different
models would be given by CR energy spectra of different elemental
groups.  Direct measurements could in principle provide such
information.  However, due to their very limited collection area
and exposure time, direct measurements across the knee energy are
practically impossible.  On the other hand, the well known
difficulties in the interpretation of EAS data have not yet
permitted such measurements either.  Thus, beyond the knee energy
very little is known about CRs other than their all-particle
energy spectrum \citep{watson98}.

The primary goal of the KASCADE experiment is to provide such
information by measuring the electromagnetic, mu\-onic, and
hadronic EAS components with high precision in each individual
event and by applying advanced methods of multi-parameter data
analyses.  Determining the energy spectrum and composition of
cosmic rays in the knee region with high precision also calls for
detailed EAS simulations, as performed by the CORSIKA package
\citep{corsika}.  Testing and improving hadronic interaction
models entering the EAS simulations is another objective of
KASCADE. As discussed below, this is realised mostly by
investigating the properties of high energy hadrons in the shower
core.  Reconstructing properties of the primary CRs from
different sets of EAS observables and comparing their results
provides another powerful means to judge the reliability of
measurements and simulations.

\begin{figure*}[t]
\includegraphics[width=17.0cm]{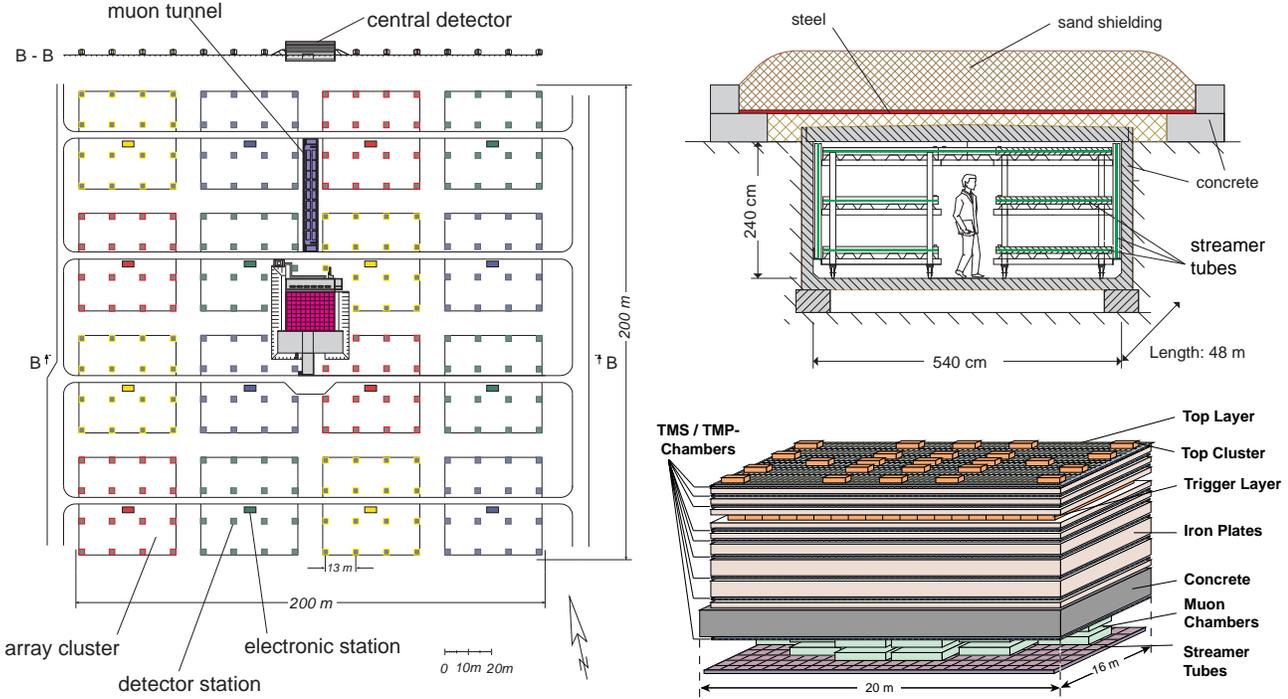}
\caption[]{Schematic layout of the KASCADE experiment (left), with 
its streamer tube tracking system (top right) and central 
detector (bottom right).}
\label{fig:kascade}
\end{figure*}

\section{Experimental}

KASCADE (\underline{Ka}rlsruhe \underline{S}hower \underline{C}ore and 
\underline{A}rray \underline{De}tector) is located at the laboratory 
site of Forschungszentrum Karls\-ruhe, Germany (at $8^{\circ}$ E, 
$49^{\circ}$ N, 110 m a.s.l.).  In brief, it consists of three major 
components (see Fig.\,\ref{fig:kascade});

\begin{enumerate}
    \item A scintillator array comprising 252 detector stations of 
    electron and muon counters arranged on a grid of $200 \times 200$ 
    m$^{2}$ and providing in total about 500 m$^2$ of $e/\gamma$- and 
    620 m$^{2}$ of $\mu$-detector coverage. The detection thresholds
    for vertical incidence are $E_{e} > 5$ MeV and $E_{\mu} > 
    230$ MeV.

    \item A central detector system (320 m$^{2}$) consisting of a
    highly-segmented hadronic calorimeter read out by 44,000
    channels of warm liquid ionization chambers distributed over
    9 read-out layers, another set of scintillation counters
    above the shielding (top cluster), a trigger plane of
    scintillation counters in the third iron gap and, at the very
    bottom, 2 layers of positional sensitive MWPC's, and a
    streamer tube layer with pad read-out for investigation of
    the muon component at $E_{\mu} > 2.4$ GeV.

    \item A $48 \times 5.4$ m$^{2}$ tunnel housing three
    horizontal and a two vertical layers of positional sensitive
    limited streamer tubes for muon tracking at $E_{\mu} > 0.8$
    GeV.

\end{enumerate}

More details about the experiment can be found in
\citet{kascade-90} and in \citet{kascade-97c}.  First correlated
data have been taken with some parts of the experiment since 1996
and with its full set-up since 2000.  At present, more than 500
Mio.\ events have been collected in a very stable mode and with a
trigger threshold of the array corresponding to $E \sim 4 \cdot
10^{14}$ eV.

\section{Tests of High-Energy Hadronic Interaction Models}

The observation of EAS provides an opportunity to study global
properties of hadronic interactions in an energy range not
accessible to man-made accelerators.  For example, the cms-energy
at the Tevatron collider corresponds to a fixed target energy in
the nucleon-nucleon system of $E_{p} \simeq 1.7 \cdot
10^{15}$\,eV. Even more importantly, the diffractive particle
production dominating the energy flux in the forward region
influences the EAS development most strongly, but has been
studied experimentally only at comparatively low energies of
$\sqrt{s}\simeq 10$ GeV \citep{kaidalov79}.  Most of the beam
energy in present collider experiments remains unobserved.  For
example, the UA5 experiment could register up to 30\,\% of the
total collision energy at $\sqrt{s}=0.9$ TeV, while the CDF
detector registers only about 5\,\% at $\sqrt{s}=1.8$ TeV. Hence,
hadronic interaction models applied to higher cms-energies and to
particle production in the very forward region rely on
extrapolations and may cause systematic uncertainties in
simulations of EAS. Additional uncertainties arise from
simulations of $p$-nucleus and nucleus-nucleus collisions
including a possible formation of a quark-gluon plasma.  Again,
such data are important for EAS interpretations but have been
studied only at low energies in the past (SPS and ISR at CERN). 
Only very recently, RHIC data at $\sqrt{s} = 200$ GeV have become
available and will be very helpful in this respect.

In KASCADE, hadronic interaction models have been tested by EAS
data employing two basic approaches: At lower primary CR
energies, where data from direct measurements still exist, the
energy spectra and rate of single hadrons not accompanied by air
showers were studied.  At higher CR energies, on the other hand,
various features of the hadronic shower cores were investigated. 
In both approaches, the sensitivity to details of the hadronic
interaction models arises mostly from the high energy threshold
applied to the hadron reconstruction in the calorimeter.

\begin{figure}[t]
\vspace*{2.0mm}
\includegraphics[width=8.3cm]{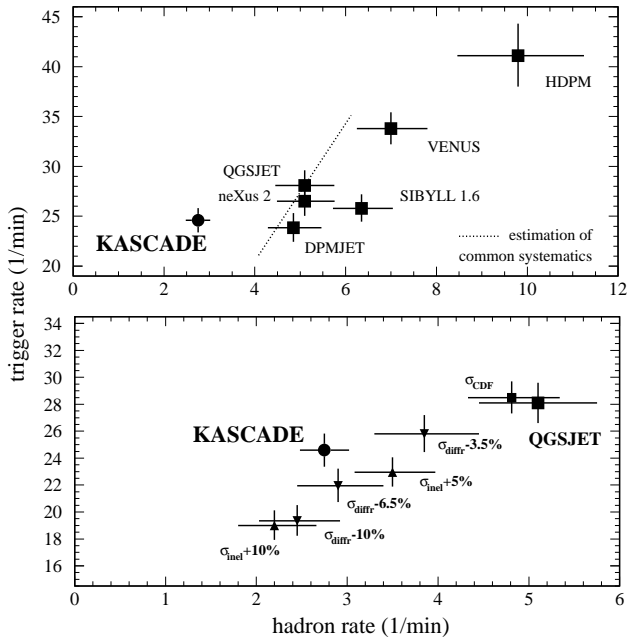}
\vspace*{-1.0mm}
\caption[]{Trigger rate vs hadron rate in the KASCADE central
detector.  The top panel compares different hadronic interaction
models to the experimental data.  The systematic uncertainty,
mostly given by the absolute flux uncertainty of the direct
experiments, is indicated by the dotted line.  The lower panel
shows results obtained from the QGSJET model with modified
inelastic cross section and diffraction dissociation
\citep{risse-thesis,kascade-01b}.}
\label{fig:trig-rates}
\end{figure}
\begin{figure}[t]
\vspace*{2.0mm}
\includegraphics[width=8.3cm]{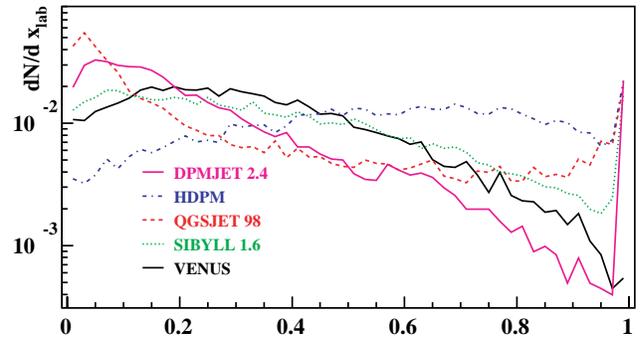}
\vspace*{-1.0mm} \caption[]{CORSIKA simulations of Feynman-$x$
distributions of leading baryons from $p+N$ interactions at
$10^{16}$ eV energy.  Five different interaction models (versions
of 1997) are shown \citep{heck-icrc01}.}
\label{fig:feynman}
\end{figure}

\subsection{Tests at Low Primary Energies: Study of Trigger Rates}

An important source of uncertainty in hadronic interaction models
originates from the energy dependent inelastic cross-sections
\citep{block00}.  Taking into account the deviations among
different experiments at Tevatron energies, the
proton-anti\-proton cross section is not known to better than
5\,\% at best \citep{avila-99}.  As shown below, these
uncertainties are amplified when predicting absolute hadron
fluxes at sea-level by means of EAS simulations.  Thus, one may
take advantage of this strong dependence and perform stringent
tests of models on the basis of EAS data.  The idea is to use
data of {\em absolute} CR fluxes up to several TeV of energy, as
obtained from balloon- and satellite borne experiments at the top
of the atmosphere.  Taking into account the measured energy
distribution and chemical composition, these particles are then
propagated through the atmosphere employing the CORSIKA
simulation package.  At the level of the KASCADE experiment, we
then ask for triggers released by either high energy hadrons
($E_{h} \ga 90$ GeV) or a minimum number of 9 muons ($E_{\mu} \ga
0.49$ GeV) detected in the central detector of KASCADE. Such an
energy cut applied to the hadrons automatically selects particles
originating from the first high-energy interactions in the
atmosphere.  Employing different hadronic interaction models,
these simulated inclusive trigger rates are then compared to
actual experimental data \citep{kascade-01b}.  Figure
\ref{fig:trig-rates} shows the results.  None of the predictions
agrees well with the experimental data, particularly the hadron
rates are overestimated by up to a factor of 3.  Furthermore,
there are also large differences found among the models.  This
convincingly proves the sensitivity of the experimental
observable to details of the interaction models.

Earlier investigations of high-energy hadrons observed in the
shower core \citep{kascade-99c} lead us to the conclusion that
QGSJET \citep{qgsjet} provides the best overall prescription of
EAS data.  This was independently concluded also from a consistency
analysis including data from different EAS experiments
\citep{erlykin98d}.  Therefore, several modifications were
applied to the QGSJET model in order to study their influence to
the predicted trigger rates.  Results are presented in the lower
panel of Fig.\,\ref{fig:trig-rates}.  Increasing the inelastic
proton-air cross section by 5\,\% (10\,\%) reduces the predicted
hadron rate by approx.\ 27\,\% (54\,\%).  Similar effects are
obtained by lowering the diffraction dissociation by up to 10\,\%
of the inelastic cross section.  Clearly, there is a strong
sensitivity to these parameters.  Since the total inelastic cross
section appears to be the better known quantity, we conclude that
the uncertainty arises mostly from the diffraction dissociation
which may be overestimated in the simulated hadron-nucleus
interactions by about 5\,\% of the inelastic cross section
$\sigma_{\rm inel}^{p{\rm +air}}$.  In fact, the trigger rates
predicted by the different models shown in
Fig.\,\ref{fig:trig-rates} (top) are strongly correlated to the
simulated shapes of the Feynman-$x$ distributions of leading
baryons.  This is demonstrated in Fig.\,\ref{fig:feynman} for
CORSIKA simulations of $p+N$ interactions at $E=10^{16}$ eV; for
the same total cross section, a large fraction of diffractive
interactions leads to high trigger rates and vice versa.  In
order to verify the aforementioned conclusions about the observed
deviations of the trigger rates it would be highly desirable to
measure the proton inelastic cross section on nitrogen and oxygen
at the highest energies at accelerators and to study the
kinematical region of diffraction dissociation.

\begin{figure}[t]
\vspace*{2.0mm}
\includegraphics[width=8.3cm]{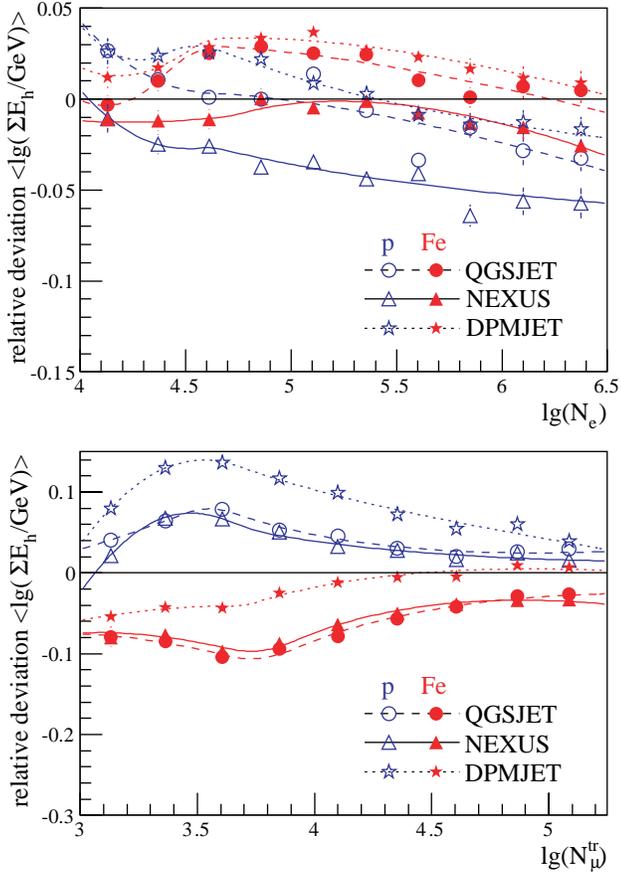}
\vspace*{-3.0mm}
\caption[]{Deviations of the simulated hadronic energy sum 
normalized to experimental KASCADE data for QGSJET, {\sc neXus 2},
and DPMJET II-5. Data in the upper and lower panel are binned as
function of electron-  and truncated muon size, respectively. 
\citep{milke02}}
\label{fig:eh-vs-ne-nmu}
\end{figure}

\subsection{Tests at High Primary Energies: Study of Hadrons in 
the Shower Core}

While the previous study selects mostly diffractive type
interactions in the atmosphere, the following analysis samples
the more 'typical' nuclear interactions leading to a fully
developed EAS. Events have to fulfil standard cuts in electron
and muon number for EAS ($N_{e} \ge 10^{4}$, $N_{\mu}^{tr} \ge
10^{3}$) and need to have at least one hadron with $E \ga
50$\,GeV reconstructed in the calorimeter.  Details of this type
of analysis were presented by \citet{milke-icrc01} at this
conference.  An example employing only the three models which are
closest to the data in Fig.\,\ref{fig:trig-rates} is depicted in
Fig.\,\ref{fig:eh-vs-ne-nmu}.  Here, the total hadronic energy
observed in the calorimeter is compared with CORSIKA simulations
for proton and iron primaries employing the QGSJET
\citep{qgsjet}, {\sc neXus 2} \citep{nexus}, and DPMJET (version
II-5) \citep{dpmjet} models.  The upper panel shows the ratio
$(\lg{\sum E_{h,sim}}-$ $\lg{\sum E_{h,data}})/\lg{\sum
E_{h,data}}$ for each of these models as a function of electron
size.  The lower panel shows the same quantity as function of the
truncated muon size, defined as $N_{\mu}^{\rm tr} = \int_{40 {\rm
m}}^{200 {\rm m}} 2\pi \rho_{\mu}(r) dr$.  Considering the steep
primary energy distribution and the different EAS developments
for light and heavy primaries, binning the data as a function of
$\lg N_{e}$ tends to enrich showers originating from light
primaries.  Therefore, proton simulations (open symbols in the
upper panel of Figure \ref{fig:eh-vs-ne-nmu}) are expected to
closely resemble the experimental data.  QGSJET simulations
follow this expectation best.  With some exceptions at lower
electron sizes, also DPMJET provides a reasonable description of
the experimental data.  However, {\sc neXus 2} clearly fails in
this test: there is too little hadronic energy at given electron
size.  Since hadrons in an EAS continuously populate the
electromagnetic component, this mismatch points to a general
problem of the balance between electromagnetic and hadronic
energy in the used version of {\sc neXus}.  However, it should be
mentioned that this version of {\sc neXus} is still considered
preliminary and will be subject of further improvements.  The
observed overall decrease of simulated/experimental hadronic
energies may be due to an increasingly heavier composition in the
experimental data at higher energies (see next section).

The truncated muon number provides an energy estimator which is
almost independent from primary mass.  Thus, repeating the
analysis and plotting the same quantities now as a function of
truncated muon size should yield results with the proton and
iron simulations being reasonably above and below the experimental
data.  Again, QGSJET exhibits the expected behaviour (lower panel
of Fig.\,\ref{fig:eh-vs-ne-nmu}).  Different from above, now the
{\sc neXus} model follows the QGSJET results very closely and
DPMJET fails the test; DPMJET II-5 produces significantly too
many hadrons so that a composition based on this model would
appear too heavy.  Such a composition contradicts
measurements from other experiments as well as those from KASCADE
(see below).  At $\lg N^{\rm tr}_{\mu} \ga 4.5$, i.e.\ energies
corresponding to $E \ga 10^{16}$ eV, a composition even heavier
than iron would be needed to describe the experimental data.  The
combined results of Fig.\,\ref{fig:eh-vs-ne-nmu} top and bottom
thus suggest that EAS simulated with DPMJET II-5 penetrate too deeply
into the atmosphere.  The conclusion is confirmed also
by studying the multiplicity and energy distribution of hadrons
in the calorimeter \citep{milke-icrc01}.

\section{Determination of the primary energy spectra and mass 
composition}

Results presented in the previous section have demonstrated that
QGSJET is still the model of choice when analysing EAS data in
the knee region.  Deviations between the Monte Carlo predictions
and experimentally observed distributions of high energy hadrons
($E_{h} \ga 90$ GeV) have become reasonably small, however,
efforts are still needed for further optimization of interaction
models.  As pointed out before, selecting high energy hadrons
amplifies the sensitivity to the actual modelling of high energy
hadronic interactions.  Low energy electrons and muons, on the
contrary, should be less affected as they result from an average
of many interactions at mostly lower cms-energies.  Thus, in the
following we will concentrate on the reconstructed electron and
truncated muon numbers.

\subsection{Unfolding Techniques}

\begin{figure*}[t]
\figbox*{}{}{\includegraphics*[width=11.0cm]{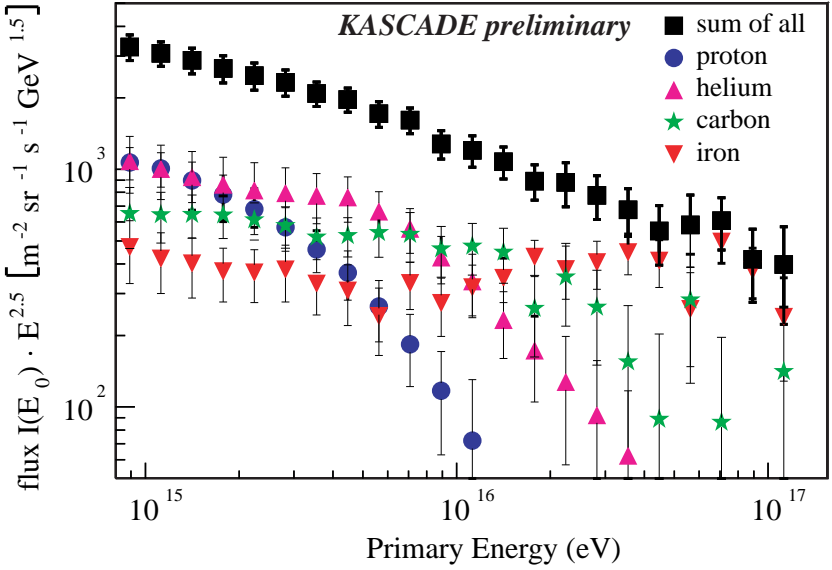}}
\vspace*{-2.0mm}
\caption[]{Preliminary energy distributions of four primary mass
groups as obtained from the unfolding procedure \citep{ulrich-icrc01}.
The sum of the 4 individual distributions represents the all-particle
CR-spectrum and is shown by the black squares. The vertical error bars
represent the statistical uncertainties which are dominated
by the statistical uncertainties of Monte Carlo simulations
(see also Fig.\,\ref{fig:e-error}).}
\label{fig:e-spec}
\end{figure*}

\begin{figure}[t]
\vspace*{2.0mm}
\includegraphics[width=8.3cm]{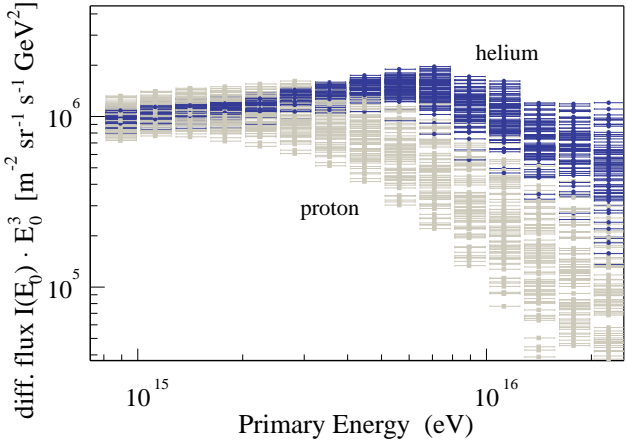}
\vspace*{-1.0mm}
\caption[]{Results of repeated unfolding procedures using 
different input parameters from the Monte Carlo simulations. The 
scattering of the symbols has been used to estimate the 
statistical uncertainty of the Monte Carlo simulation. 
See also Fig.\,\ref{fig:e-spec} and text for details
\citep{ulrich-thesis}.}
\label{fig:e-error}
\end{figure}

It is well known that for given zenith angle, i.e.\ fixed
atmospheric thickness, $N_{e}$ and $N_{\mu}^{\rm tr}$ depend
simultaneously on the mass, $A$, and energy, $E$, of the primary
particle.  Besides their specific {\em average\/} relation
$(N_{e}, N_{\mu}^{\rm tr}) \leftrightarrow (E, A)$ also their
{\em fluctuations\/} are known to change with energy and mass. 
Clearly, such effects are to be taken into account when
reconstructing the properties of the primary particles,
particularly in the presence of steeply falling energy spectra. 
Mathematically, the observed electron and muon shower size
distributions at a given zenith angle bin can then be written as
\begin{displaymath}
    \frac{dJ}{d\lg N_{e,\mu}} = \sum_{A} \!\!
    \int_{-\infty}^{\infty} \!\!\!
    \frac{dJ_{A}(\lg E)}{d\lg E} \:
    p_A(\lg N_{e,\mu} | \lg E) \: d\lg E
\end{displaymath}
where the sum runs over all primary masses $A$.  The quantity
$p_A(\lg N_{e,\mu} | \lg E)$ denotes the probability for a
primary particle of mass $A$ and energy $\lg E$ to be
reconstructed as an air shower with electron- and muon size $\lg
N_{e}$ and $\lg N_{\mu}^{\rm tr}$, respectively.  The equation thus
represents a set of Fredholm integral equations of 1$^{\rm
st}$ kind.  In practice, not all primary masses $A$ will be taken
into account, but only a representative group of different
primary particles, here protons, helium, carbon, and iron. 
Suitable methods to solve the inverse equation, i.e.\ to infer
the physical quantities of interest, $dJ_{A}/d\lg E$, from the
experimentally observed $dJ/d\lg N_{e,\mu}$ distributions, are
unfolding algorithms.  Here, we have chosen the iterative scheme
of the Gold-algorithm \citep{gold-64} which tries to minimize the
$\chi^{2}$ functional and allows only non-negative solutions
\citep{ulrich-icrc01}.

The kernel function $p_{A}$ needs to be determined from EAS and
detector simulations and has to account for all kinds of physical
and experimental effects, such as fluctuations of shower sizes,
trigger and detection efficiencies or reconstruction
accuracies.  It has been determined from high statistics
CORSIKA/QGSJET simulations generated with thinning option for
3 different bins of zenith angle ($0^{\circ}$-$18^{\circ}$,
$18^{\circ}$-$25.8^{\circ}$, $25.9^{\circ}$-$32.3^{\circ}$), four
primary masses and for a large number of fixed primary energies. 
The individual $N_{e}$ and $N_{\mu}^{tr}$ distributions obtained
for each primary energy and mass were then parametrized and the
variation of the parameters with shower size, zenith angle,
primary energy, and primary mass was used in the analysis. 
Details of this method are described in \citet{ulrich-icrc01} and
by \citet{ulrich-thesis}.

\begin{figure*}[t]
\vspace*{2.0mm}
\figbox*{}{}{\includegraphics*[width=11.0cm]{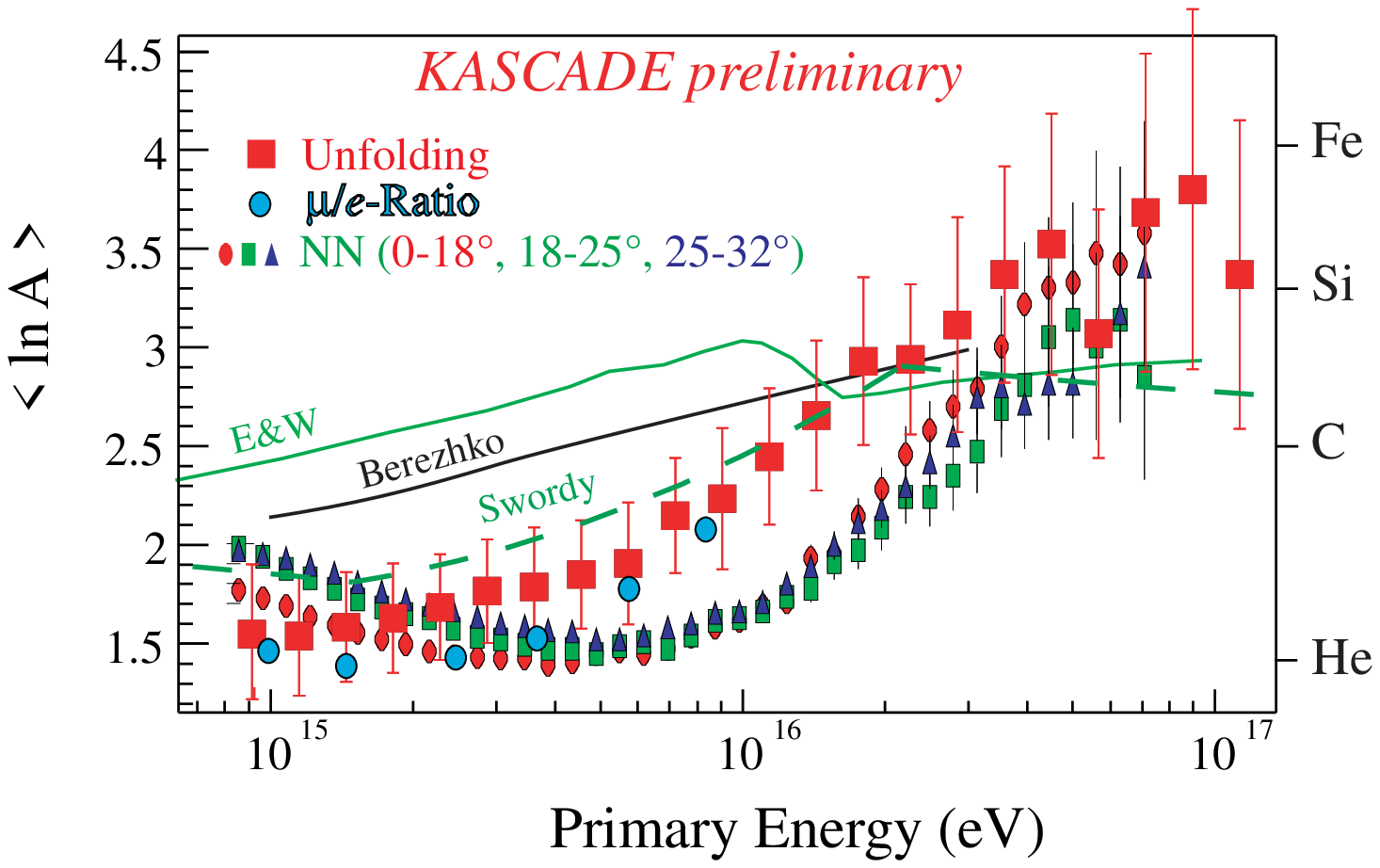}}
\vspace*{-1.0mm}
\caption[]{Mean logarithmic mass as a function of the primary 
energy. The theoretical calculations are adopted from \citet{erlykin97a} 
(E\&W), \citet{berezhko99}, and \citet{swordy-95}. The KASCADE 
data (symbols) are based on the unfolding procedure, an earlier 
analysis of the $\lg N_{\mu}/\lg N_{e}$ ratio \citep{weber-99}, 
and are from a neural network algorithm applied to the reconstructed
electron and muon numbers at 3 different zenith angels 
\citep{roth-icrc01}.}
\label{fig:ln-a}
\end{figure*}

Results based on the experimental $\lg N_{e}$ and $\lg
N_{\mu}^{\rm tr}$ shower distributions of all 3 zenith angle bins
are presented in Fig.\,\ref{fig:e-spec} for 4 primary mass
groups.  Each of the reconstructed energy distributions shows a
knee like structure which is shifted towards higher energies with
increasing mass.  The knee in the all-particle spectrum at about
4 PeV appears to me mostly caused by the break of the light
elements proton and helium.  This is well in agreement to earlier
analyses of KASCADE data \citep{glasstetter-99,kascade-02a}.

It should be stressed that no assumption about spectral shapes or
mass abundances is made in the analysis presented here, only the
measured electron and muon shower size distributions at 3
zenith angles and the simulated response function $p_{A}$ are
used.  The error bars in Fig.\,\ref{fig:e-spec} represent
statistical errors only and are dominated by the limited
statistics of the presently available CORSIKA simulations. 
Statistical errors of the experimental data correspond to about
20 months of data taking and are very small compared to that of
the simulations.

In order to investigate the robustness of the reconstructed
energy spectra, the unfolding algorithm has been repeated many
times employing different sets of input parameters to the kernel
function $p_{A}$.  These parameters describing the shapes of the
simulated $N_{e}$ and $N_{\mu}^{tr}$ distributions for fixed
primary energies (see above) have been randomly distributed
according to their individual fit uncertainties assuming Gaussian
error shapes.  The result of this study is shown in
Fig.\,\ref{fig:e-error} for the reconstructed proton and helium
energy spectra.  The shoal of symbols, all obtained using the
same experimental data, clearly demonstrates the presence of a
knee at distinct energies in both of the reconstructed spectra. 
A closer inspection of the spectra below the knee energy exhibits
a steeper proton spectrum as compared to helium and a smoother
turnover into a power law behaviour above the knee.  However, it
should be kept in mind that the reconstructed proton spectrum is
significantly affected by the large shower fluctuations limiting
the experimental energy resolution.  In fact, simulations have
shown that even a sharp break would result in a smoothness
similar to the observed one \citep{ulrich-thesis}.  The observed
scattering of the reconstructed fluxes at a given energy has been
used to estimate the statistical uncertainties of the unfolding
procedure.  Adding this in quadrature to the experimental
statistical errors yields the values shown in
figure\,\ref{fig:e-spec}.  Clearly, improving the statistics of
the Monte Carlo simulations in the near future will reduce this
error.

The still preliminary energy spectra support the picture of a
rigidity dependent scaling of the knee position.  However, since
$Z/A \simeq 0.5$ for all shown mass groups with $A \ge 4$, the crucial
test for distinguishing between a $E/A$ and $E/Z$ scaling of the
knee is given by the comparison of the proton and helium spectra. 
As can be seen from Fig.\,\ref{fig:e-error}, the knee between $p$
and He is shifted by factor of $\approx 2.5$ favouring the
rigidity picture. However, because of the different spectral
shapes, such a comparison is not straightforward and further
investigations are needed to verify that conclusion.  Also, the
sensitivity to the used hadronic interaction model needs to be
studied.  This may shift the overall energy scale and flux of the
individual curves but is anticipated to have only minor influence
to the relative positions of the observed knees.

\subsection{Chemical Composition and Comparison to Neural Network
Approaches}

Individual energy spectra of the type shown above are ideally
suited for comparisons with cosmic ray source- and acceleration
models.  However, up to now most EAS experiments have only been
able to extract the all-particle energy spectrum and the mean
logarithmic mass, $\langle \ln A \rangle$.  As a consequence,
most theoretical papers present their results just in terms of
these rather insensitive and inclusive variables.  Thus, for
completeness we plot in Fig.\,\ref{fig:ln-a} the quantity
$\langle \ln A \rangle$ as a function of the primary energy.  As
expected from the structure of the individual energy spectra, we
find an increasingly heavier composition at energies above the
knee.  The data are compared to theoretical calculations by
\citet{erlykin97a}, \citet{berezhko99}, and \citet{swordy-95}. 
Also, other experimental results from KASCADE are included. 
Common to all models and data is an increasingly heavier mass at
energies above the knee.  Only the single source model of
\citet{erlykin97a} shows some structure at $ \sim 10^{16}$ eV due to
the assumed occurrence of an iron peak.  As can be seen from that
figure, the models differ significantly in their prediction of
the mean mass and, except for the Swordy calculation, predict
heavier masses than extracted from the experimental data.

In addition to the results from the unfolding procedure, we
include earlier results from an analysis of the $\lg
N_{\mu}/\lg N_{e}$ ratio \citep{weber-99} and from a neural
network algorithm applied to the reconstructed electron and muon
numbers at 3 different zenith angels \citep{roth-icrc01}. 
Within their errors, we find similar results from the unfolding
procedure and from the $\lg N_{\mu}/\lg N_{e}$ ratios.  However, the
neural network approach yields a similar trend only at energies
above the knee.  In contrast to the other results, the
composition shows a weak decrease in the mean logarithmic mass by
about 0.5 units in the energy range from about 1\,PeV to 4\,PeV.
Similar trends have been observed also by Cherenkov air shower
measurements.  (For a compilation of several data sets see
\citet{swordy-02}).  At present, it cannot be said whether the
effect is real or an artefact of the data analysis, such as due
to improperly accounted EAS fluctuations, neural network training
effects, etc.\,.  For example, it is easy to realize that the
occurrence of a break in the individual energy spectra combined
with the well known stronger shower fluctuations for light
primaries could qualitatively account for such an effect. 
Generally, this would produce a bias towards a lighter
composition the steeper the power law distribution of the true
proton spectrum is.

The experimental data match well to direct measurements recently
reported by the JACEE and RUNJOB experiments up to $E \simeq
10^{15}$ eV \citep{jacee-98,runjob-01}.  While RUNJOB reports an
almost constant composition of $\langle \ln A \rangle \simeq 1.6$
in the energy range $10^{13}$ - $10^{15}$ eV, that of JACEE tends
to increase from 1.5 to 2.5 in the same energy range.  This may
point to some methodical problems in at least one of the two
experiments.  However, the statistical and systematic
uncertainties of both experiments are still sufficiently large at
$10^{15}$ eV ($\langle \ln A \rangle \simeq 1.5^{+2.0}_{-1}$ for
RUNJOB and $2.5 \pm 0.6$ for JACEE) to be compatible to one
another and to the reported KASCADE data.  Even though only
40\,\% of the available RUNJOB data have been included in that
analysis and some more data being available from JACEE, the
unsatisfactory situation of lacking statistics in direct
experiments around $10^{15}$ eV will hardly change in the near
future.

\section{Discussion and Outlook}

We have presented results on tests of the latest versions of
currently available hadronic interaction models implemented into
CORSIKA. Only three of the models (QGSJET, DPMJET II-5, and {\sc
neXus 2}) could describe the experimentally observed trigger rates
at a reasonable level.  The models Sibyll 1.6, VENUS, and HDPM
overestimate the hadron trigger rates by up to a factor of 3. 
The differences in the predicted trigger rates are strongly
correlated to the amount of diffractive dissociation, as can be
seen from the Feynman-$x$ distributions of leading baryons in
Fig.\,\ref{fig:feynman}.  Clearly, there is a need to improve the
knowledge about diffractive interactions up to the highest
energies by conducting either dedicated experiments or by
increasing the forward acceptance of currently operating and
planned accelerator experiments.

Investigating the properties of QGSJET, {\sc neXus}, and DPM\-JET
in EAS at energies around the knee reveals some problems also for
the latter two models.  {\sc neXus 2} predicts too little hadronic
energy at given electron size, i.e.\ the balance between
electromagnetic and hadronic energy disagrees with experimental
data.  DPMJET II-5, on the other hand, produces too many hadrons (and
electrons) at given muon size, i.e.\ the EAS penetrate too deeply
into the atmosphere.  Thus, applying DPMJET II-5 for composition
studies would mimic a too heavy composition.  Combining all
results, QGSJET still provides the best overall description of
EAS in the knee region.  Nevertheless, a continuation of such
tests up to higher primary energies with optimized experimental
observables and with tuned cuts to the hadron energy, appears
mandatory in absence of adequate accelerator experiments.

Analysing the energy spectrum and composition of CRs in the knee
region calls for suitable analysis techniques which properly
account for mass and energy dependent EAS fluctuations and
which enable the simultaneous usage of many (correlated) EAS
parameters.  Here, we have presented results from an `inclusive'
unfolding technique applied to the electron and muon shower size
distributions at different zenith angles as well as results from
an `exclusive' neural network algorithm.  The notations
`inclusive' and `exclusive' refer to the fact that the unfolding
algorithm does not aim at providing information about the primary
energy and mass in single events, but only for the full sample of
events.  In other words, the technique is applied to measured
{\em shower size distributions\/} and it supplies {\em energy
distributions}.  This lack of information in single events
appears to be well compensated for by the sensitivity to details
of the primary energy spectra.  Employing this approach provides,
for the first time, primary energy spectra for different
mass species.  Without making any `a priory' assumption about the
individual energy spectra, we find power-law distributions with a
knee-like structure in each of the distributions.  The observed
shift of the knee energies supports the picture of a rigidity
dependent cut-off, as expected from most astrophysical models of
CR origin and acceleration.

The neural network approach confirms these overall trends but
quantitative differences to the unfolding technique are observed
and deserve further studies.  Using only the electron and muon
shower size as input does not allow to adequately distinguish
more than two or at most three mass groups on a shower-by-shower
basis.  However, another advantage besides the shower-by-shower
information is the simple inclusion of more EAS observables and
the automatic consideration of correlations among the parameters. 
Such multi-parameter analyses have revealed systematic effects in
the reconstructed energy and mass depending of the EAS
observables used \citep{kascade-02b}, thereby demonstrating the
need for further improvements of the EAS simulations.

\begin{figure}[t]
\vspace*{2.0mm}
\includegraphics[width=8.3cm]{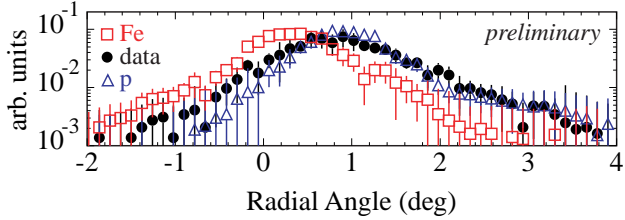}
\vspace*{-1.0mm}
\caption[]{Distribution of the muon production height, expressed
in terms of the radial angle for $3.75 \le \lg N_{\mu}^{\rm tr}
\le 4.0$.  The data are compared to CORSIKA/QGSJET simulations of
proton and iron primaries \citep{buettner-icrc01}.}
\label{fig:mu-height}
\end{figure}

Clearly, much more work is still needed to verify and manifest
the important finding of the observed break in the individual
energy spectra.  Despite the enormous progress made in recent
years, parallel efforts are still required to (i) improve the
modelling of EAS and to understand also the tails in the
distributions of EAS observables, (ii) to continue developing
advanced and complementary EAS analysis techniques, (iii) to
provide additional EAS observables by the experimental data, and
(iv) to improve the statistics and reconstruction accuracies
particularly at the highest energies ($\sim 10^{17}$ eV) in order
to verify the occurrence of an iron knee.  All of these aspects
are presently addressed by the KASCADE collaboration.  For
example, besides improvements to CORSIKA and its interaction
models, a new two-dimensional non-parametric unfolding algorithm
has been developed to account also for the shower-by-shower
correlations of the electron and muon numbers
\citep{roth-02,kascade-02c}.  The preliminary results obtained
with that method agree well with those presented in this article. 
Also, new observables just become available in KASCADE and will
be included in future analyses.  As an example we show in
Fig.\,\ref{fig:mu-height} preliminary results of the muon
production height as extracted from the data of the streamer tube
tracking detector in the tunnel north of the central detector
\citep{buettner-icrc01}.  This observable will become very
powerful particularly at energies $\ga 10^{16}$ eV, but further
optimization of the tracking algorithm and cut parameters is
still needed before including the observable in the standard set
of energy and composition parameters.  Finally, in order to
improve the measurements of primary energy and mass up to
energies of about $10^{18}$ eV, the KASCADE and EAS-TOP
experiment have been combined at the site of Forschungszentrum
Karls\-ruhe to form a new experiment, called KASCADE-Grande
\citep{bertaina-icrc01}.  After finishing the deployment of 37
stations of 10 m$^{2}$ detection area each over an area of $\sim
0.5$ km$^{2}$ by the end 2001 the new experiment will start to
take data in spring 2002.

\begin{acknowledgements}
This work has been supported by Forschungs\-zen\-trum Karlsruhe,
by the Ministry for Research of the Federal Government of Germany
(BMBF), by a grant of the Romanian National Agency for Science,
Research and Technology, by a research grant of the Armenian
Government, and by the ISTC project A116.  The collaborating
group of the Cosmic Ray Division of the Soltan Institute of
Nuclear Studies in Lodz is supported by the Polish State
Committee for Scientific Research.  The KASCADE collaboration
work is embedded in the frame of scientific-technical cooperation
(WTZ) projects between Germany and Armenia, Poland, and Romania.
\end{acknowledgements}

\end{document}